\documentclass[pra, showpacs, floatfix, preprint, endfloats*]{revtex4}
\usepackage{amsmath}
\usepackage{amssymb}
\usepackage{revsymb}
\usepackage{graphicx}
\usepackage{mathrsfs}
\begin{document}
\title{Laser photon merging in proton-laser collisions}
\author{A. \surname{Di Piazza}}
\email{dipiazza@mpi-hd.mpg.de}\author{K. Z. \surname{Hatsagortsyan}}
\email{k.hatsagortsyan@mpi-hd.mpg.de}
\author{C. H. \surname{Keitel}}
\email{keitel@mpi-hd.mpg.de}
\affiliation{Max-Planck-Institut f\"ur Kernphysik,
    Saupfercheckweg 1, D-69117 Heidelberg, Germany}

\date{\today}

\begin{abstract}
The quantum electrodynamical vacuum polarization effects arising in the collision of a high-energy proton beam and a strong, linearly polarized laser field are investigated. The probability that laser photons merge into one photon by interacting with the proton`s electromagnetic field is calculated taking into account the laser field exactly. Asymptotics of the probability are then derived according to different experimental setups suitable for detecting perturbative and nonperturbative vacuum polarization effects. The experimentally most feasible setup involves the use of a strong optical laser field. It is shown that in this case measurements of the polarization of the outgoing photon and and of its angular distribution provide promising tools to detect these effects for the first time.

 
\end{abstract}
 
\maketitle

\section{Introduction}
With the advent of quantum field theory our idea of vacuum has been strongly modified with respect to that of classical physics. Vacuum is no more intended as a literal nothing but as a physical entity with a complex structure. Within the framework of Quantum Electrodynamics (QED), the interest in investigating the properties of quantum vacuum has been growing considerably in the last years, mostly due to the advancement of the strong field laser technique. In particular, many experimental schemes have been put forward to measure these properties (see the recent Reviews \cite{Reviews} and the references therein). In fact, QED predicts that in the presence of very intense electromagnetic fields quantum vacuum manifests dielectric properties due to the polarization of the virtual electron-positron background and even breaks down in stronger fields when virtual electron-positron pairs, acquiring sufficient energy from the field, become real \cite{Dittrich_b_1985,Greiner_b_1985,Dittrich_b_2000}. The strength scale of the electromagnetic field where the QED vacuum effects become apparent is determined by the so-called critical fields: $E_{cr}=m^2/e=1.3\times 10^{16}\;\text{V/cm}$ and $B_{cr}=m^2/e=4.4\times 10^{13}\;\text{G}$ (here $-e<0$ and $m$ are the electron charge and mass, respectively and natural units with $\hbar=c=1$ are used). The investigation of the properties of QED vacuum is closely connected with the possibility of testing QED in the presence of strong background electromagnetic fields. However, the values of the critical fields are far beyond the strength of electromagnetic fields that can be produced in terrestrial laboratories. A remarkable exception to this fact are highly charged nuclei. In the context of QED, ``highly charged nuclei'' are nuclei with a charge number $Z$ such that $Z\alpha\lesssim 1$, with $\alpha=e^2/4\pi\approx 1/137$ the fine structure constant. In fact, the electric field of highly charged nuclei at the typical QED length $\lambda_c=1/m$ (Compton length) is comparable with the critical field $E_{cr}$. This explains the recent experimental successes in measuring the Delbr\"{u}ck scattering \cite{Delbexp}, i. e. the scattering of a high-energy photon by the field of a highly charged ion, and the photon splitting in a Coulomb field \cite{Splitexp} confirming the theoretical predictions \cite{LMS,LMMST03}.

The development of laser technology is opening up the possibility of probing the properties of QED vacuum and of testing QED in the presence of intense wave fields. Although the next generation of petawatt laser systems is expected to reach an intensity of ``only'' $10^{22}\text{-}10^{23}\;\text{W/cm$^2$}$ \cite{Norby_2005}, different theoretical schemes have been put forward to approach the so-called Schwinger intensity of $I_{cr}=E_{cr}^2/8\pi=2.3\times 10^{29}\;\text{W/cm$^2$}$ \cite{ZW}. A significant step towards the Schwinger intensity will be realized in the near future at the Extreme Light Infrastructure (ELI) where unprecedented intensities of the order of $10^{25}\text{-}10^{26}\;\text{W/cm$^2$}$ will be attained \cite{Laser_ELI}. The ever-increasing available laser intensities have stimulated numerous theoretical proposals to observe ``refractive'' vacuum polarization effects (VPEs) induced by strong laser fields \cite{VPEs1,VPEs2,VPEs3,VPEs4,VPEs5,VPEs6,VPEs7,VPEs8,VPEs9,VPEs10,VPEs11,VPEs12,VPEs13,Milstein_2005,Di_Piazza_2007_a}, while the observation of ``absorptive'' VPEs in colliding laser beams, namely electron-positron pair creation, requires field intensities of the order of $I_{cr}$ \cite{Burke_1997}. 
However, laser-assisted pair production was observed about ten years ago at the Stanford Linear Accelerator Center (SLAC) in the collision of a high-energy electron beam ($\approx 50\;\text{GeV}$) and an optical laser beam with an intensity of the order of $10^{19}\;\text{W/cm$^2$}$ \cite{Burke_1997}. Due to the large electron energy, the effective laser field amplitude and frequency in the rest frame of the electron are much larger than their value in the laboratory frame and this has allowed electron-positron pair creation.

A QED process in a strong laser field, that is characterized by a typical four-momentum of a particle $p^{\mu}$ and by a laser field with four-vector potential amplitude $A_0^{\mu}$, field tensor amplitude $F_0^{\mu\nu}$, electric field amplitude $E_0$ and frequency $\omega_0$,
depends on the two invariant parameters $\xi=e\sqrt{-A_{0,\mu}A_0^{\mu}}/m=eE_0/m\omega_0$ and $\chi=e\sqrt{-(F_{0,\mu \nu}p^{\nu})^2}/m^3$ (the metric $g^{\mu\nu}=\text{diag}\{+1,-1,-1,-1\}$ is employed) \cite{Ritus_review}. The classical field parameter $\xi$ can be interpreted as the work done by the laser field on the electron in 
the Compton length $\lambda_c=1/m$ in units of the laser photon energy and, therefore, determines the role of multiphoton processes (at $\xi \gtrsim 1$ the process is multiphoton). The same parameter $\xi$ also determines the adiabaticity of the process (the limit $\xi \gg 1$ and $\chi$ fixed corresponds to the constant crossed field case) \cite{Keldysh}. The quantum field parameter $\chi$ determines the magnitude of the quantum nonlinear effects. 
When $\xi \ll 1$ multiphoton effects play a minor role and the probability of the process goes to its perturbative weak field limit described by the Feynman diagram(s) of lowest order in the laser field. In the opposite case of $\xi \gg 1$ the process is adiabatic, the laser field is almost constant and the photon picture does not hold anymore. In this limit another perturbative weak field regime exists when the parameter $\chi$ is much smaller than unity. In the following, by ``perturbative regime'' we mean that either $\xi\ll 1$ or $\chi\ll 1$ when the process probability can be represented as a power expansion in the corresponding small parameter. In the case of two equal, counterpropagating laser beams, the parameter $\chi$ results from the $F_0^{\mu\nu}$ of one laser beam and from the $p^{\nu}$ of a photon of the other laser beam: $\chi\sim(\omega_0/m)(E_0/E_{cr})$. The schemes proposed in \cite{VPEs1,VPEs2,VPEs3,VPEs4,VPEs5,VPEs6,VPEs7,VPEs8,VPEs9,VPEs10,VPEs11,VPEs12,VPEs13,Milstein_2005} to observe VPEs in colliding laser beams are limited to \emph{perturbative} processes. This is because for usual parameters of strong optical lasers, though the condition $\xi\gg 1$ is easily fulfilled, nevertheless, $\chi \ll 1$ since $\omega_0 \ll m$ and $E_0\ll E_{cr}$, and, consequently, the nonperturbative effects are suppressed \cite{Photon_splitting}. 
However, the VPEs can be increased if a strong laser beam collides with a high-energy proton beam. Analogously to the SLAC experiment \cite{Burke_1997}, this is due to the Lorentz boost of the laser field amplitude and frequency in the rest frame of the relativistically fast moving proton. The perturbative 
regime ($\chi \ll 1$ and $\eta\equiv\chi/\xi\ll 1$) of scattering of a strong laser field with arbitrary polarization by a Coulomb field has been considered in \cite{Milstein_2005}. The first attempt to go beyond the perturbative regime of VPEs in this process has been done in the early work of Ref. \cite{Yakovlev_1967}, where the laser photon merging was considered in the scattering of a strong \emph{circularly} polarized laser field by a Coulomb field. Due to the symmetry of the system, however, only two laser photons can merge 
if the laser field is circularly polarized, i. e. the possibility of multiphoton 
VPEs is in principle ruled out in this case. Moreover, the author explicitly 
considers only regions of parameters where nonperturbative effects either 
are completely negligible or cancel out each other, in such a way that the final 
results are, in fact, perturbative. Already in Ref. \cite{Yakovlev_1967} it was pointed out that the large Lorentz boost of a high-energy proton beam enhances the cross-section of the photon merging process in the perturbative case. Moreover, the enhancement of the laser field amplitude and frequency 
may open a way to nonperturbative regimes: $\xi\gg 1$ and $\chi \gtrsim 1$.  In \cite{Di_Piazza_2008} we have done the first calculation of probabilities of nonperturbative refractive VPEs in the collision of a high-energy proton beam with a linearly polarized, strong laser field and proposed a setup that will allow an observation of the merging of multiple pairs of laser photons into a single high-energy photon due to VPE induced by the proton field and the laser field.

The proton features fit very well the requirements of this setup. In fact, on the one hand it is light enough to be accelerated to very high energies like up to $980\;\text{GeV}$ at the Tevatron or even up to $7\;\text{TeV}$ at the forthcoming Large Hadron Collider (LHC) \cite{PDG}, such that the laser field amplitude and frequency in the proton rest frame are significantly larger than their values in the laboratory frame. On the other hand, the proton is heavy enough that multiphoton Thomson scattering of the laser photons by the proton itself is strongly suppressed. This is of great advantage with respect to employing, for example, an electron beam because, as we will see, multiphoton Thomson scattering represents a background of our process of laser photon merging. 

The nonperturbative VPEs during a proton and a laser beam collision manifest themselves in two ways. First, the scaling of the photon merging probability is very sensitive to the laser field parameters. Thus, in the limit $\xi\gg 1$ the probability $P_{2n}$ of the merging of $2n$ photons scales as $P_{2n} \sim \chi^{4/3}$ if $\chi\gg 1$, while in the perturbative case ($\chi\ll 1$) the scaling is $P_{2n} \sim \chi^{4n}$. Second, the multiphoton processes (merging of $2n$ laser photons with $n>1$), while being negligibly small in the perturbative regime, become significant and observable in the nonperturbative regime. In our previous Letter \cite{Di_Piazza_2008} we have investigated the most favorable case for observation of multiphoton VPEs by employing proton accelerators already available and the next generation of petawatt optical laser systems. Accordingly, we have calculated the probabilities of the photon merging in the domain of $\xi \gg 1$ and $\chi$ fixed. However, there are other interesting regimes of parameters which show peculiarities in the amplitudes of the process and which are connected with the possibility of the experimental observation of laser photon merging with  the next generation of x-ray-free electron lasers (X-FELs) like those available in the near future at DESY \cite{Tesla} and at SLAC \cite{LCLS}. The investigation of these regimes is carried out in the present paper.

In the present paper we thoroughly investigate the laser photon merging process during the proton and laser beam collision in nonperturbative regimes, taking into account exactly the influence of the laser field. Starting from the general expressions of the amplitudes of the process, 
we obtain analytical asymptotics valid in different parameters regions and analyze ranges of parameters not considered yet. In particular, additionally to the previous results, we obtain new analytical expressions for the amplitudes of the process in the domain of $\xi \ll 1$ and for any $\eta$, as well in the regime of $\eta \gg 1$ and $\xi \gtrsim 1$ that are relevant for X-FEL. We also analyze the experimental feasibility of the process with different possible experimental devices. Attention is devoted to the question of how advantageous for the photon merging process could be the complementary virtues of an X-FEL with respect to an optical laser, namely, the high photon energy, the high repetition rate and the relatively large space-time volume of the laser beam. As we will see, the use of optical strong laser fields is much more favorable from an experimental point of view. For this reason, in addition to the results in \cite{Di_Piazza_2008}, we also investigate in this case the polarization and the complete angular distribution of the emitted photons.

The paper is organized as follows: the theoretical model is considered in Sec. II. The asymptotics of the probability of the process are analyzed in Sec. III. In the same section, in each asymptotic limit the corresponding physical situations are considered and the process feasibility is discussed. The summary of Sec. IV concludes the paper. For the sake of completeness, we have reported in an Appendix the expression of the polarization operator found in \cite{Baier_1976} that will be our starting point here.
%
%
\section{Theoretical model}
\begin{figure}
\begin{center}
\includegraphics[width=\textwidth]{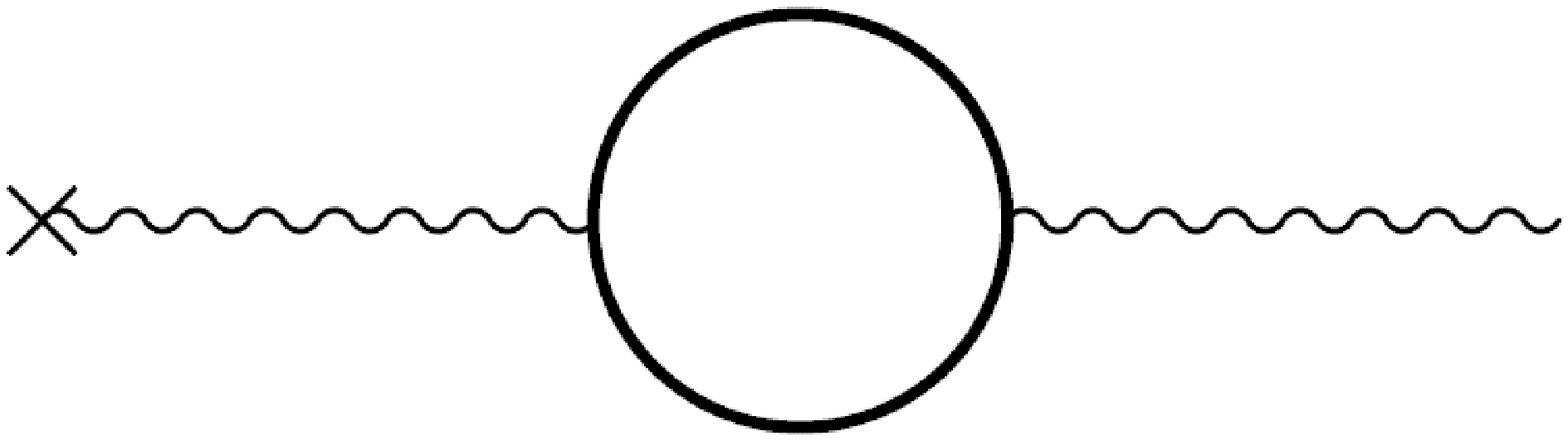}
\end{center}
\caption{Feynman diagram corresponding to the process of laser photon merging induced by the VPEs in a proton field and in a strong laser field. The thick fermion lines indicate that the dressed propagator in a laser field is employed.}
\label{Feynman_diagram}
\end{figure}

Below, we consider the VPEs arising in the head-on collision of a high-energy proton with a strong laser beam. We assume that the laser beam propagates in the positive $y$ direction and that it is linearly polarized along the $z$ direction. The amplitude and the intensity of the laser field will be indicated as $E_0$ and $I_0=E_0^2/8\pi$, respectively, whereas $\omega_0$ is the laser frequency. Concerning the proton, we have modeled it as an external field due to its large energy and we have ensured that the recoil on the proton of the emitted high-energy photon was negligible in the numerical examples. We assume that the proton moves with velocity $\beta$ along the negative $y$ direction. The proton mass is indicated as  $M=938\;\text{MeV}$ and its relativistic Lorentz factor as $\gamma=1/\sqrt{1-\beta^2}$. As the proton is considered as an external field, the Feynman diagram representing the process of photon merging is that shown in Fig. \ref{Feynman_diagram}. The crossed photon leg corresponds to the proton electromagnetic field which is represented by the four-vector potential
\begin{equation}
\label{A}
A^{\mu}(t,\mathbf{r})=\frac{e}{4\pi}\frac{u^{\mu}}{\sqrt{x^2+\gamma^2(y+\beta t)^2+z^2}},
\end{equation}
with $u^{\mu}=(\gamma,0,-\beta\gamma,0)$ being the four-velocity of the proton. Since the proton charge number is $Z=1$, the proton field can be safely taken into account perturbatively at lowest order. The thick fermion lines in Fig. \ref{Feynman_diagram} are dressed fermion propagators which exactly include the laser field. This is achieved by employing the Volkov states to build the fermion propagator \cite{Landau_b_4_1982}. Finally, the photon line on the right in Fig. \ref{Feynman_diagram} indicates the emitted photon resulting from the merging. We indicate its four-momentum as $k^{\mu}=(\omega,\mathbf{k})$, with $\omega=|\mathbf{k}|$ being the outgoing photon energy, and its two polarization four-vectors as $e_a^{\mu}(k)$, with $a\in \{1,2\}$. The presence of the laser field can in principle induce a mass different from zero of the outgoing photon through vacuum polarization and the appearance of a third, longitudinal photon mode \cite{Baier_1976}. In order to estimate these effects we assume that the emitted photon counterpropagates with the laser field. From the expression of the polarization operator in a laser field one sees that at fixed $\xi\lesssim 1$ then these effects are $\alpha$ times smaller than those considered here, apart from the asymptotic case $\eta'\equiv 2\omega_0\omega/m^2\to\infty$ in which they are $\alpha\log\eta'$ times smaller \cite{Ritus_review}. In both cases, then, they can be safely neglected for realistic parameters. Instead, in the constant crossed field limit $\xi\gg 1$ and $\chi'=2(\omega/m)(E_0/E_{cr})$ fixed, these effects scale as $\alpha\chi^{\prime\, 2}$ if $\chi'\ll 1$ and $\alpha\chi^{\prime\,2/3}$ if $\chi'\gg 1$ \cite{Ritus_review}. In this situation they become important at $\chi'\gtrsim 10^3$ but we will not consider this extreme case.

The two amplitudes of the laser photon merging process depending on the final photon polarization state are given by (see Fig. \ref{Feynman_diagram})
\begin{equation}
M_a(k)=\int \frac{d^4q}{(2\pi)^4} A_{\mu}(q)\Pi^{\mu\nu}(q,k)\frac{e^*_{a,\nu}(k)}{\sqrt{2\omega}},
\end{equation}
where $\Pi^{\mu\nu}(q,k)$ is the vacuum polarization tensor in a laser field \cite{Baier_1976} and where
\begin{equation}
\label{A_q}
A^{\mu}(q)=2\pi e\frac{u^{\mu}}{q_x^2+q_z^2+\left(\frac{q_0}{\beta\gamma}\right)^2}\delta((qu))
\end{equation}
is the four-dimensional Fourier transform of the four-vector potential in Eq. (\ref{A}), with $(qu)=\gamma(q_0+\beta q_y)$ and $\delta(x)$ being the Dirac $\delta$ function. The vacuum polarization tensor $\Pi^{\mu\nu}(q,k)$ in a laser field has been calculated in \cite{Baier_1976} and we report in the Appendix the final expression valid for an elliptically polarized laser field with the main axes directed along the $x$ and $z$ direction \textit{and for initial and final four-momenta $q^{\mu}$ and $k^{\mu}$ such that, as in the situation at hand, $q^2<0$ and $k^2=0$}. By starting from that expression of the vacuum polarization tensor in the case of a linearly polarized laser field along $z$ (one has to set $A_{10}=E_0/\omega_0$ and $A_{20}=0$ in Eq. (\ref{A_mu})), it can be shown that the amplitudes $M_a(k)$ can be written as
\begin{equation}
\label{M}
M_a(k)=\frac{m^2e\alpha}{i\gamma}\sum_{n=1}^{\infty}\frac{\delta(\beta(2n\omega_0-k_y)+2n\omega_0-\omega)}{2n\omega_0(\omega-k_y)\sqrt{2\omega}}\sum_{j=1}^2c_{j,2n}(u\Lambda_j)(e_a^*\Lambda_j).
\end{equation}
To obtain this equation we have exploited the four-dimensional $\delta$ function present in Eq. (\ref{C_r}) that gives $q^{\mu}+2nk_0^{\mu}=k^{\mu}$ and then $q^2=-2n(k_0k)=-2n\omega_0(\omega-k_y)$. Also, we have employed the auxiliary four-vectors $\Lambda_j^{\mu}$ given in Eq. (\ref{Lambda_j_app}):
\begin{equation}
\label{Lambda_j}
\Lambda_j^{\mu}=a_j^{\mu}-\frac{(ka_j)}{(k_0k)}k_0^{\mu},
\end{equation}
where $a_1^{\mu}=(0,0,0,1)$, $a_2^{\mu}=(0,1,0,0)$ and where $k_0^{\mu}=(\omega_0,0,\omega_0,0)$ is the four-momentum of the laser photons, and we have introduced the coefficients
\begin{equation}
\label{c_j_n_ex}
\begin{split}
c_{j,2n}&=i^n\int_{-1}^1dv\int_0^{\infty}\frac{d\rho}{\rho}e^{-i\Phi_{2n}}\left\{\xi^2\left[\frac{A}{2}\mathcal{J}^*_n(z_{2n})-\sin^2\rho J_n(z_{2n})\right]\delta_{j,1}\right.\\
&\qquad\qquad\left.+\xi^2\frac{\sin^2\rho}{1-v^2}\mathcal{J}_n(z_{2n})+\frac{\eta_{2n}}{2}\left(n-i\frac{1-v^2}{4\rho}\right)J_n(z_{2n})\right\},
\end{split}
\end{equation}
where $\delta_{j,j'}$ is the Kronecker $\delta$ function, $\Phi_{2n}=2n\rho+4\rho\{1+\xi^2[1-\sin^2(\rho)/\rho^2]/2\}/\eta_{2n}(1-v^2)$, $z_{2n}=2\rho\xi^2[\sin^2(\rho)/\rho^2-\sin (2\rho)/2\rho]/\eta_{2n}(1-v^2)$, $A=1+\sin^2(\rho)/\rho^2-\sin (2\rho)/\rho$ and $\mathcal{J}_n(z)=J_n(z)+iJ'_n(z)$, with $J_n(z)$ being the ordinary Bessel function of order $n$ and $J'_n(z)$ its derivative. As it is clear from the expressions of the coefficients $c_{j,2n}$, the invariant amplitudes of the laser photon merging process essentially depend on the laser field only through the two gauge- and Lorentz-invariant parameters
\begin{align}
\label{xi}
\xi&\equiv\frac{eE_0}{m\omega_0},\\
\label{eta}
\eta_{2n}&\equiv\frac{(k_0k_{2n})}{m^2}=2n\frac{\omega_0^2}{m^2}\frac{1+\beta}{1+\beta\cos\vartheta}(1-\cos\vartheta).
\end{align}
In the expression of the parameter $\eta_{2n}$, we have introduced the four-momentum $k^{\mu}_{2n}$ of the outgoing photon when $2n$ ($n\ge 1$) laser photons merge, which is defined as
\begin{equation}
k^{\mu}_{2n}=\omega_{2n}(1,\sin\vartheta\sin\varphi,\cos\vartheta,\sin\vartheta\cos\varphi).
\end{equation}
Here, $\omega_{2n}$ is the energy of the outgoing photon when $2n$ laser photons merge:
\begin{equation}
\label{omega_2n}
\omega_{2n}=2n\omega_0\frac{1+\beta}{1+\beta\cos\vartheta},
\end{equation}
while $\vartheta$ and $\varphi$ are the polar and the azimuthal angle in the momentum space, being the laser propagation direction ($y$ axis) the polar axis. A few observations are in order. First, as it can also be inferred from the Furry theorem, only an even number of laser photons can merge in the process at hand \cite{Landau_b_4_1982}. Second, we note that due to gauge and Lorentz invariance the scalar coefficients $c_{j,2n}$ cannot depend on the scalar products $(k_{2n}a_j)$ and this is why they do not depend as well on the azimuthal angle $\varphi$. Third, the parameter $\chi$ defined in the Introduction depends here on the number of laser photons merged and it is equal to
\begin{equation}
\label{chi_n}
\chi_{2n}\equiv\xi\eta_{2n}=\sqrt{\frac{I_0}{I_{cr}}}\frac{2n(1+\beta)\omega_0}{m}\frac{1-\cos\vartheta}{1+\beta\cos\vartheta}.
\end{equation}
This is because the momentum $p^{\mu}$ entering the definition of $\chi$ is given here by $k^{\mu}_{2n}$ or, which is equivalent, by the momentum transfer from the proton. It is worth mentioning here that, in general, the series in Eq. (\ref{M}) also contains terms with $n\le 0$. These terms vanish in the situation at hand because, \textit{as we have shown in the Appendix,} the four-dimensional $\delta$ functions in Eq. (\ref{C_r}) cannot be fulfilled. However, on the one hand, the term with $n=0$ contributes to the modification to the Coulomb potential of the proton induced by vacuum polarization \cite{Milstein_2005}. On the other hand, it also gives rise to a radiative correction to Thomson scattering without net absorption of laser photons. Both these corrections are $\alpha$ times smaller than the effect considered here and will be neglected. The coefficients $c_{j,2n}$ show a highly nonperturbative dependence on the laser field parameters through $\xi$ and $\eta_{2n}$. The parameter $\xi$, as already mentioned in the Introduction, is responsible for the magnitude of high-order vacuum effects in the laser intensity. The quantity $\sqrt{2\eta_{2n}}$ is the total energy of a laser photon and the outgoing photon in their center-of-momentum frame when $2n$ laser photons merge in unit of the electron mass.

At this point, in order to calculate the total rate $\mathcal{R}$ of emitted photons one has to employ the amplitudes $M_a(k)$ in Eq. (\ref{M}) and apply the Fermi golden rule \cite{Landau_b_4_1982}:
\begin{equation}
\label{R}
\begin{split}
\mathcal{R}&\equiv\sum_{a=1}^2\int\frac{d\mathbf{k}}{(2\pi)^3}|M_a(k)|^2\\
&=\frac{\alpha^3}{16\pi^2}\frac{m^4}{(1+\beta)\gamma^2\omega_0^3}\sum_{n=1}^{\infty}\frac{1}{n^3}\int\frac{d\Omega}{4\pi}\frac{1}{(1-\cos\vartheta)^2}\sum_{a=1}^2\left\vert\sum_{j=1}^2c_{j,2n}(u\Lambda_j)(e_a^*\Lambda_j)\right\vert^2.
\end{split}
\end{equation}
Here, we have assumed a space-time integration volume equal to unity and we have exploited the Dirac $\delta$ function to perform the integral on the energy of the outgoing photon. $d\Omega$ indicates the differential solid angle relative to the outgoing photon direction. By employing the expressions of the four-vectors $\Lambda_j$ in Eq. (\ref{Lambda_j}), one can easily show that
\begin{equation}
(u\Lambda_j)=\gamma(1+\beta)\frac{\sin\vartheta}{1-\cos\vartheta}(\cos\varphi\,\delta_{j,1}+\sin\varphi\,\delta_{j,2}).
\end{equation}
Also, since the four-vectors $\Lambda_j^{\mu}$ are both perpendicular to $k^{\mu}$ (see Eq. (\ref{Lambda_j})), when we sum in Eq. (\ref{R}) over the outgoing photon polarizations we can employ the substitution \cite{Mandl_b_1984}
\begin{equation}
\sum_{a=1}^{2}e_a^{\mu}e_a^{*\,\nu}\longrightarrow -g^{\mu\nu}.
\end{equation}
Note that, in general, the sum should include the four polarization states of the photon (in this case the above equation becomes an equality). However, due to gauge invariance only the two transverse photon polarization states contribute to the photon merging rate. This can also be verified through a direct calculation by observing that $(k\Lambda_j)=0$ and by reminding that $k^2=0$. In this way, the integration over the angle $\varphi$ in Eq. (\ref{R}) can be easily taken and, by introducing the rate $\mathcal{R}_{2n}$ of the laser photon merging with an absorption of $2n$ laser photons, we obtain
\begin{equation}
\label{dRdtheta_n}
\frac{d\mathcal{R}_{2n}}{d\vartheta}=\frac{\alpha^3}{64\pi^2}\frac{(1+\beta)m^4}{\omega_0^3}\frac{\sin^3\vartheta}{(1-\cos\vartheta)^4}\frac{|c_{1,2n}|^2+|c_{2,2n}|^2}{n^3}.
\end{equation}
From this expression it is clear that the evaluation of the laser photon merging rate reduces to the evaluation of the two coefficients $c_{j,2n}$. Their general expressions in Eq. (\ref{c_j_n_ex}) are rather complex and, as we will see in the next section, they can be simplified in different asymptotical regions of the parametric space $\xi\text{-}\eta_{2n}$.

\begin{figure}
\begin{center}
\includegraphics[width=\textwidth]{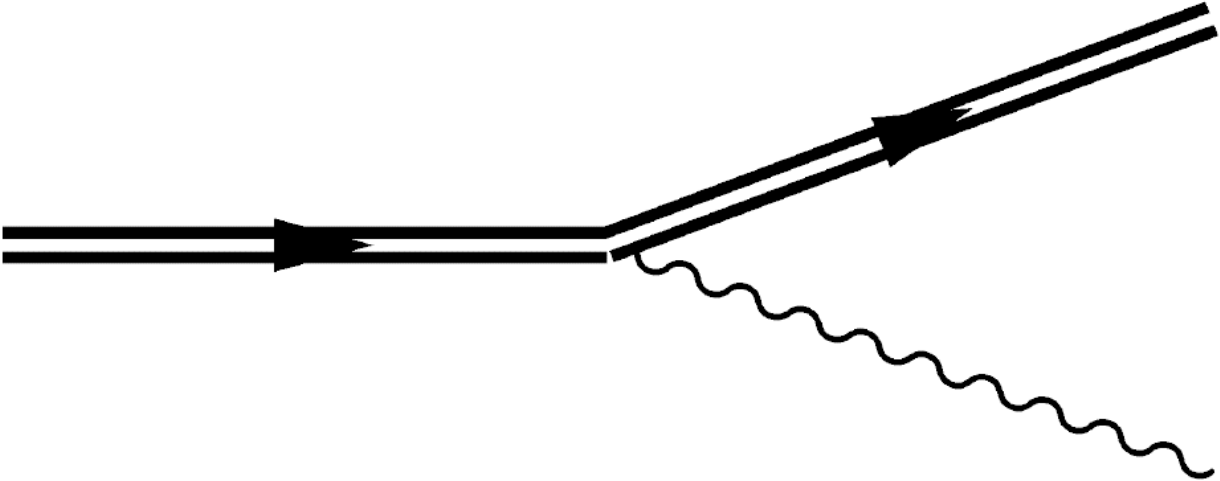}
\end{center}
\caption{Feynman diagram corresponding to the process of multiphoton Thomson scattering by a proton (thick double line) in a laser field.}
\label{Multiphot_Thomson}
\end{figure}

We conclude this section with the following observation that we have mentioned in the Introduction. It can easily be seen that the photons emitted via 2$n$-multiphoton Thomson scattering of laser photons by the proton have the same energy as the photons emitted via the 2$n$-laser photon merging process because the kinematics of the two processes is the same (see the review \cite{Ritus_review} for a detailed discussion of multiphoton Thomson scattering in the case of electrons). Both processes occur in the collision of a proton and a laser beam and they interfere. As it is clear from the Feynman diagram in Fig. \ref{Feynman_diagram}, the photon merging process is qualitatively different from the multiphoton Thomson scattering (see Fig. \ref{Multiphot_Thomson}). In the latter, the photons of the laser interact \emph{directly} with the proton through its electric charge. In the former, instead, the laser photons interact with the \emph{electromagnetic field} of the proton through a virtual electron-positron pair. An immediate consequence of this difference is that the process of multiphoton Thomson scattering, though being a process of lower order in $\alpha$ than the process of $2n$-laser photon merging, is nevertheless suppressed due to the large proton mass $M$ and in the limit $M\to \infty$ the amplitude of the process vanishes. This is not the case of the laser photon merging process which has a finite leading-order amplitude in the limit $M\to \infty$ that exactly corresponds to the Feynman diagram in Fig. \ref{Feynman_diagram}.
%
%
\section{Asymptotics of the coefficients $c_{j,2n}$}

The expressions (\ref{c_j_n_ex}) of the coefficients $c_{j,2n}$ hold for any value of the two physical parameters $\xi$ and $\eta_{2n}$. However, these expressions are rather complex and difficult to use for quantitative estimations. Simpler analytical expressions can be found in different limits of the parameters $\xi$ and $\eta_{2n}$. As we will see below, this corresponds to various physical situations when different laser systems could be employed to observe the process of laser photon merging like, for example, X-FELs or strong optical lasers. To the sake of completeness, we report here also the results obtained in \cite{Di_Piazza_2008} concerning the case of an optical laser field (see Par. III.B). The expressions of the amplitudes for the case in which the strong laser field is an X-FEL are obtained here for the first time and are reported in Par. III.A and Par. III.C.
%
%
\subsection{The case of an X-FEL ($\xi\ll 1$, $\eta_{2n}$ fixed)}
\label{small_xi}
We consider in this paragraph the asymptotic expressions of the coefficients $c_{j,2n}$ when the parameter $\xi$ can be considered much smaller than unity, while the parameter $\eta_{2n}$ is arbitrary but fixed. This case is particularly suitable for treating the practical situation in which the strong laser field is that of an X-FEL. In fact, due to the large laser frequency and the low intensity, the parameter $\xi$ for an X-FEL is much smaller than unity. As a check, we consider the following laser parameters that will be available with the Tesla project at DESY \cite{Tesla}: a laser frequency $\omega_0=3.1\;\text{keV}$, a laser peak power of $80\;\text{GW}$ and a spot radius of $30\;\text{$\mu$m}$ corresponding to a peak intensity $I_0=2.8\times 10^{15}\;\text{W/cm$^2$}$, and we merely obtain $\xi=2\times 10^{-5}$. On the other hand, being the laser frequency only two-three orders of magnitude smaller than the electron mass, it would then be limiting (as we will see in a numerical example below) to restrict anyway the values of the parameter $\eta_{2n}$ (see Eq. (\ref{eta})) mostly in the case of a high-energy proton beam.

\begin{figure}
\begin{center}
\includegraphics[width=\textwidth]{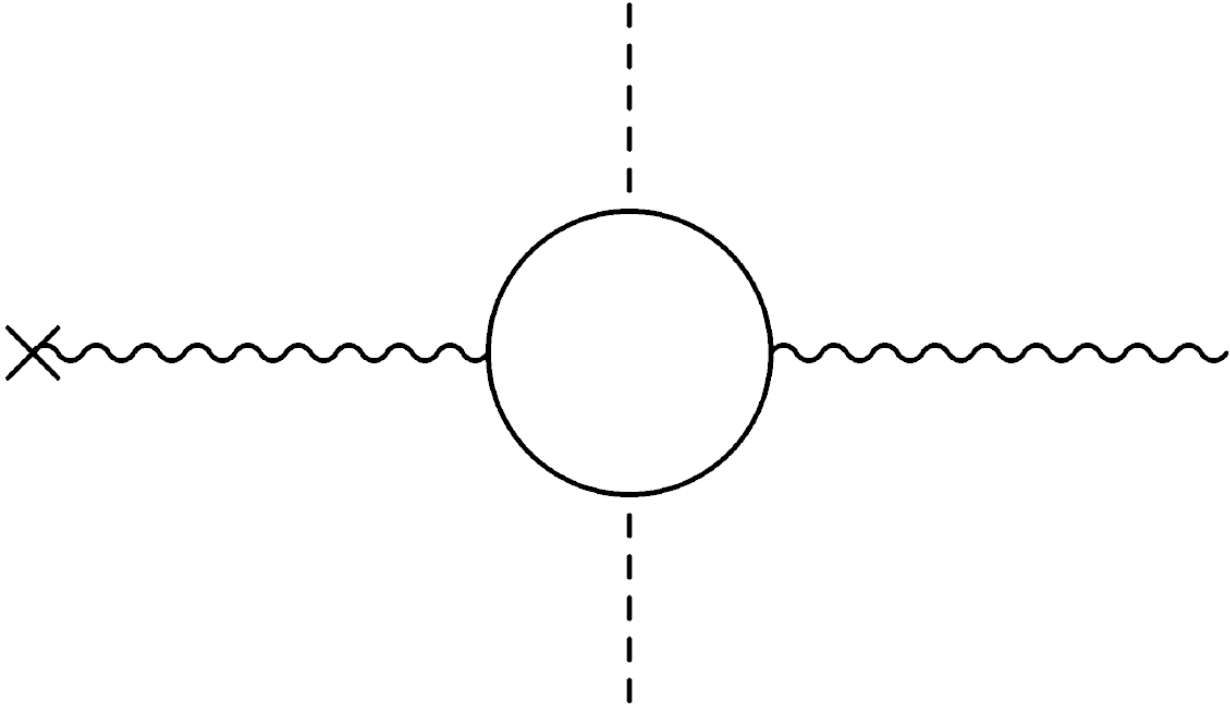}
\end{center}
\caption{One of the Feynman box-diagrams corresponding to the process of 2-laser photon (dashed lines) merging in a proton field. In the remaining two, the laser photon lines are both either above or below the proton field and the outgoing photon lines.}
\label{Box_diagram}
\end{figure}

As we have mentioned, in the present case with $\xi\ll 1$, the laser field can be treated perturbatively. By expanding the exponential and the Bessel functions in the coefficients $c_{j,2n}$ with respect to the small parameter $\xi$, one sees that the leading term of the series is proportional to $\xi^{2n}$. This term corresponds to the process of laser photon merging in which $2n$ laser photons merge. The next terms in the expansion result proportional to $\xi^{2(n+l)}$, with $l$ being a positive integer, and correspond to the process of $2n$-laser photon merging accompanied by the exchange of $l$ laser photons without \emph{net} absorption. Due to the dependence on $\xi^{2n}$, higher-order processes are suppressed and the leading contribution to the laser photon merging process comes from the term with $n=1$ and $l=0$, i. e. from the merging of two laser photons without any additional exchange. One of the Feynman box-diagrams representing this process is shown in Fig. \ref{Box_diagram}. \textit{In the remaining two, the laser photon lines are both either above or below the proton field and the outgoing photon lines. The use of the Volkov states as electron states automatically takes also into account the contribution of all these diagrams. It is also worth observing here that another diagram is present with four external photon lines, namely the one with one laser photon and two proton field lines. This diagram is the lowest-order diagram contributing to the Delbr\"uck scattering of a laser photon by the proton. The probability of this process is roughly $\alpha^2/\xi^2$ times the probability of 2-laser photon merging and then it can be dominating in the present case. However, the two processes are experimentally distinguishable because the energy of the photon emitted via Delbr\"uck scattering is half that of the photon emitted via 2-laser photon merging.} 

After expanding the exponential and the Bessel functions in Eq. (\ref{c_j_n_ex}), the integral over $\rho$ can be performed (we call to mind that the convergence of this integral as $\rho\to\infty$ is ensured by the usual $m^2\to m^2-i\varepsilon$ prescription) and we obtain:
\begin{equation}
\label{c_j_n_xi_small}
\begin{split}
c_{j,2}&=\frac{\xi^2}{32}\int_0^1\frac{dv}{1-v^2}\left\{4[4-g(1-v^2)]\text{arctanh}\left(\frac{2}{g}\right)+(-1)^j4(1-v^2)\right.\\
&\qquad\qquad\left.-[8(v^{2(2-j)}-g)+3^{2-j}g^2(1-v^2)]\log\left(1-\frac{4}{g^2}\right)\right\},
\end{split}
\end{equation}
where we have introduced the quantity $g=2+4/\eta_2(1-v^2)$. The remaining integration in Eq. (\ref{c_j_n_xi_small}) can be performed with Mathematica and after a number of simplifications (we have employed, in particular, the properties of the Euler's dilogarithm function described in \cite{Bateman_b_1953}) the final result is
\begin{align}
\label{c_1_2}
\begin{split}
c_{1,2}&=\frac{\xi^2}{8\eta_2^2}\left\{-\eta_2^2+12(1-\eta_2)\sqrt{\eta_2}\left(\sqrt{1+\eta_2}\text{arcsinh}\sqrt{\eta_2}-\sqrt{2+\eta_2}\text{arcsinh}\sqrt{\frac{\eta_2}{2}}\right)\right.\\
&\qquad\qquad\left.-2\left[(3-\eta_2)(1+2\eta_2)\text{arcsinh}^2\sqrt{\eta_2}-2(3+4\eta_2-\eta_2^2)\text{arcsinh}^2\sqrt{\frac{\eta_2}{2}}\right]\right\},
\end{split}\\
\label{c_2_2}
\begin{split}
c_{2,2}&=\frac{\xi^2}{8\eta_2^2}\left\{\eta_2^2+4(1-3\eta_2)\sqrt{\eta_2}\left(\sqrt{1+\eta_2}\text{arcsinh}\sqrt{\eta_2}-\sqrt{2+\eta_2}\text{arcsinh}\sqrt{\frac{\eta_2}{2}}\right)\right.\\
&\qquad\qquad\left.-2(1-\eta_2)\left[(1+2\eta_2)\text{arcsinh}^2\sqrt{\eta_2}-2(1+\eta_2)\text{arcsinh}^2\sqrt{\frac{\eta_2}{2}}\right]\right\}.
\end{split}
\end{align}
It is remarkable that these expressions are exact in the parameter $\eta_2$ and they are explicitly expressed only through elementary functions. Moreover, the analytical structure of the amplitudes in the present case is completely different from that in the case of a strong optical laser field (see Eq. (\ref{c_n}) below). In particular, altough the present regime is perturbative in the sense that $\xi\ll 1$, a complex dependence of the amplitudes is observed on the laser frequency $\omega_0$ through the parameter $\eta_2$.

In the limit of $\eta_2$ much smaller than unity (small laser frequencies), we obtain from Eqs. (\ref{c_1_2})-(\ref{c_2_2})
\begin{equation}
c_{j,2}=\left(\frac{7}{4}\right)^{j-1}\frac{\xi^2\eta_2^2}{45}.
\end{equation}
In this case the total rate $\mathcal{R}_2$ can be easily calculated and is given by
\begin{equation}
\label{R_2}
\mathcal{R}_2=\frac{13\alpha^3}{19440\pi^2}\omega_0\gamma^4(1+\beta)^5\left(\frac{I_0}{I_{cr}}\right)^2.
\end{equation}
This result does not agree with the corresponding one calculated in \cite{Milstein_2005} and presented there in Eq. (61) where a factor four is missing. We point out, however, that all the conclusions of the paper \cite{Milstein_2005} are unaffected by this misprint. The result in Eq. (\ref{R_2}) coincides with the rate calculated by starting from the lowest-order Euler-Heisenberg Lagrangian density \cite{Dittrich_b_2000}
\begin{equation}
\label{L}
\mathscr{L}=\frac{1}{2}(E^2-B^2)+\frac{2\alpha^2}{45m^4}\left[(E^2-B^2)^2+7(\mathbf{E}\cdot\mathbf{B})^2\right],
\end{equation}
in which the total electromagnetic field $(\mathbf{E},\mathbf{B})$ is replaced by the sum of the proton field, the laser field and the photon field and it is expanded up to linear terms in the proton and the photon fields. This result which may appear obvious when also looking at Fig. \ref{Box_diagram} deserves investigation. For example, the same would not be true if a laser photon line were substituted in Fig. \ref{Box_diagram} by a proton line (Delbr\"{u}ck scattering) \cite{Papatzacos_1975}. In fact, the use of the Euler-Heisenberg Lagrangian density is allowed when the momenta flowing in the electron propagators (see Fig. \ref{Box_diagram}) are much smaller than the electron mass $m$. In the case of Delbr\"{u}ck scattering this is not true for the momentum flowing between the two photon legs representing the proton electromagnetic field. Instead, in our case the momentum $q^{\mu}$ absorbed by the proton is fixed by the energy-momentum conservation relation $q^{\mu}+2k_0^{\mu}=k_2^{\mu}$. In this way, $q^2=-4(k_0k_2)=-4\eta_2 m^2$ and $|q^2|\ll m^2$, all the momenta flowing in the box diagram are then much smaller than $m$.

In the opposite limit $\eta_2\gg 1$, one obtains for the coefficients in Eqs. (\ref{c_1_2}) and (\ref{c_2_2}) the following expressions:
\begin{equation}
\label{c_j_n_xi_small_large_eta}
c_{j,2}=\frac{\xi^2}{8}\left[3(\log 2-1)^2-\frac{4}{j}+2\log 2\,\log\eta_2\right].
\end{equation}
The behavior of the amplitudes in this limit depends on the polarization of the laser field: in \cite{Yakovlev_1967} it was found that in the same limit the amplitudes of the laser photon merging process in a circularly polarized laser field become independent of the parameter $\eta_2$. Although interesting from a theoretical point of view, this case is of less practical interest because, in general, it would require too high laser frequencies. In fact, if the emission angle $\vartheta$ is much smaller than unity, the condition $\eta_2\gg 1$ means 
that $\omega_0\vartheta/m\gg 1$. Instead, at intermediate angles such that $\cos\vartheta\lesssim 1$, the condition $\eta_2\gg 1$ implies that $\omega_0\gg m$. A less restrictive condition is obtained if $\vartheta\approx \pi$ (outgoing photon almost collinear with the proton). In fact, in this case the condition $\eta_2\gg 1$ is fulfilled if $\gamma(1+\beta)\omega_0/m\gg 1$. This strong inequality can in turn be fulfilled by combining an X-FEL with $\omega_0\sim 1\text{-}10\;\text{keV}$ \cite{Tesla,LCLS} with a large accelerator like Tevatron or LHC for which $\gamma\sim 10^3$ \cite{PDG}. However, one has to bear in mind that at $\vartheta$ exactly equal to $\pi$ the laser photon merging rate vanishes.

\begin{figure}
\begin{center}
\includegraphics[width=\textwidth]{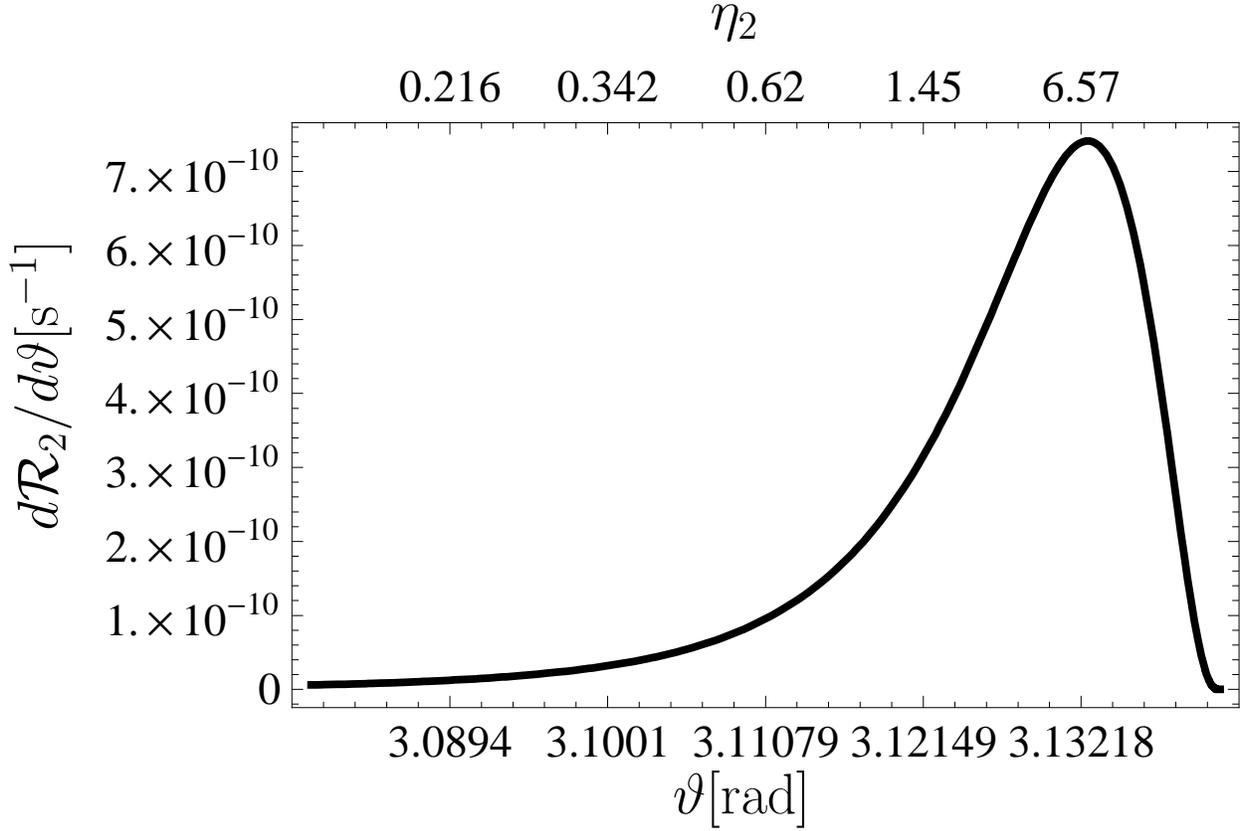}
\end{center}
\caption{Differential 2-photon merging rate as a function of the emission angle $\vartheta$. The parameters of the laser field are $\omega_0=3.1\;\text{keV}$ and $I_0=2.8\times 10^{15}\;\text{W/cm$^2$}$ and the proton's relativistic Lorentz factor is $\gamma=10^3$. The upper horizontal axis shows the values of the parameter $\eta_2$ as a function of $\vartheta$.}
\label{Small_xi}
\end{figure}

We conclude this paragraph by showing in Fig. \ref{Small_xi} the rate of 2-photon merging by employing the aforementioned laser parameters that will be available at the Tesla facility at DESY: $\omega_0=3.1\;\text{keV}$ and $I_0=2.8\times 10^{15}\;\text{W/cm$^2$}$. The figure shows the behavior of the rate, calculated by employing Eqs. (\ref{c_1_2}) and (\ref{c_2_2}), as a function of the emission angle $\vartheta$ at a typical value $\gamma=10^3$ for the Lorentz relativistic factor of the proton. The rate is almost peaked for back-emitted photons where the parameter $\eta_2$ is larger. At those values of $\vartheta$ where the spectrum has a peak, one can see that the value of $\eta_2$ can be larger than unity. Finally, the figure also shows the advantage of using strong optical laser fields in order to observe the laser photon merging process experimentally: in the present case, the photon merging rate is several orders of magnitude smaller than that obtained in the next section for the optical laser case (cf. also the numerical simulation relative to Fig. 2 in \cite{Di_Piazza_2008}). It can be easily checked that the high effective repetition rate of the X-FEL (30000 pulses in one second \cite{Tesla}) and its relatively large space-time volume ($50\;\text{$\mu$m}$ waist size and $100\;\text{fs}$ pulse duration \cite{Tesla}) cannot compensate for the negligibly small rate.
%
%
\subsection{The case of a strong optical laser field ($\xi\gg 1$, $\eta_{2n}\ll 1$ and $\xi\eta_{2n}$ fixed)}
\subsubsection{Total rate}

Available optical lasers easily exceed the electron relativistic threshold $\xi\approx 1$. In fact, for a typical optical photon energy of $\omega_0=1\;\text{eV}$, the condition $\xi=1$ is fulfilled at a laser intensity of about $10^{18}\;\text{W/cm$^2$}$. On the other hand, as the laser photon energy is much smaller than the electron mass $m=0.5\;\text{MeV}$, one can also assume that $\eta_{2n}\ll 1$, even in the case when
the gamma factor of the proton is as large as about $7000$ like at the LHC. Now, in the asymptotic limit $\xi\to\infty$ and $\eta_{2n}\to 0$ in such a way that the parameter $\chi_{2n}=\xi\eta_{2n}$ remains finite (see Eq. (\ref{chi_n})), it can be shown that the leading-order contributions to the coefficients $c_{j,2n}$ are
\begin{equation}
\label{c_n}
c_{j,2n}=e^{-i\pi/3}\int_0^1dv\int_0^{\infty}\frac{d\lambda}{\lambda}e^{-\exp(i\pi/3)\lambda-x_{2n}}\left\{j\chi_{2n}^2\lambda^2\frac{1-v^{4/j}}{16}[I_n(x_{2n})-I'_n(x_{2n})]+\frac{I_n(x_{2n})}{\lambda}\right\}
\end{equation}
where $x_{2n}=\chi_{2n}^2\lambda^3(1-v^2)^2/96$ and where $I_n(x)$ is the modified Bessel function of order $n$ and $I'_n(x)$ its derivative. In this limit the coefficients $c_{j,2n}$ depend only on the parameter $\chi_{2n}$. The above expressions are valid for any value of $\chi_{2n}$ and we have also calculated the two limiting expressions of $c_{j,2n}$ when the parameter $\chi_{2n}$ is much smaller than unity and much larger than unity (see Eqs. (6) and (7) in \cite{Di_Piazza_2008}). In the former case, the rate of photon merging with absorption of more than two laser photons is negligible and the total photon merging rate with absorption of two laser photons is given again by Eq. (\ref{R_2}). This is an interesting point that deserves examination. In fact, as we have seen in the previous paragraph, the perturbative approach holds when $\xi\ll 1$ while we are working in the opposite limit $\xi\gg 1$ here. However, one can show that for the process at hand the limit $\xi\gg 1$ and $\chi_{2n}\ll 1$ gives the same result of the limit $\xi\ll 1$ and $\eta_{2n}\ll 1$. This is a peculiarity of the photon merging process and it was also pointed out in \cite{Yakovlev_1967} in the case of a circularly polarized laser field: it was noted by Affleck and Kruglyak in \cite{VPEs4} that this is not true in the case of photon splitting in a laser field (see also \cite{Di_Piazza_2007_a}) where the results of the two limits are different. In the case of photon splitting $\chi=\xi\eta$, where $\eta=(k_0k)/m^2$, with $k^{\mu}$ being the four-momentum of the incoming photon. In the limit $\xi\gg 1$ and $\chi\ll 1$ the photon splitting amplitude corresponds to the hexagon Feynman diagram with three laser photon legs and not to the box diagram as in the limit $\xi\ll 1$ and $\eta\ll 1$.

The regime of the photon merging $\xi \gg 1$ and $\chi_{2n} \gtrsim 1$ is the most interesting nonperturbative regime. In the limit $\chi_{2n}\gg 1$ the scaling of the probability of photon merging is nonperturbative (see Eq. (7) in \cite{Di_Piazza_2008}): $c_{j,2n}\sim \chi_{2n}^{2/3}$. In this regime merging of many laser photon pairs is not negligible as in the perturbative regime ($\chi_{2n}\ll 1$) where it is damped by a small factor $\chi_{2n}^{2n}$. Nevertheless, if $\xi,\chi_{2n} \gg 1$, then the rate of multiphoton processes decreases slowly with increasing $n$ as $1/n^5$. In \cite{Di_Piazza_2008} we have analyzed the feasibility of observing experimentally nonperturbative VPEs in a strong laser field in the regime $\xi \gg 1$ and $\chi_{2n} \gtrsim 1$. 

\begin{figure}
\begin{center}
\includegraphics[width=\textwidth]{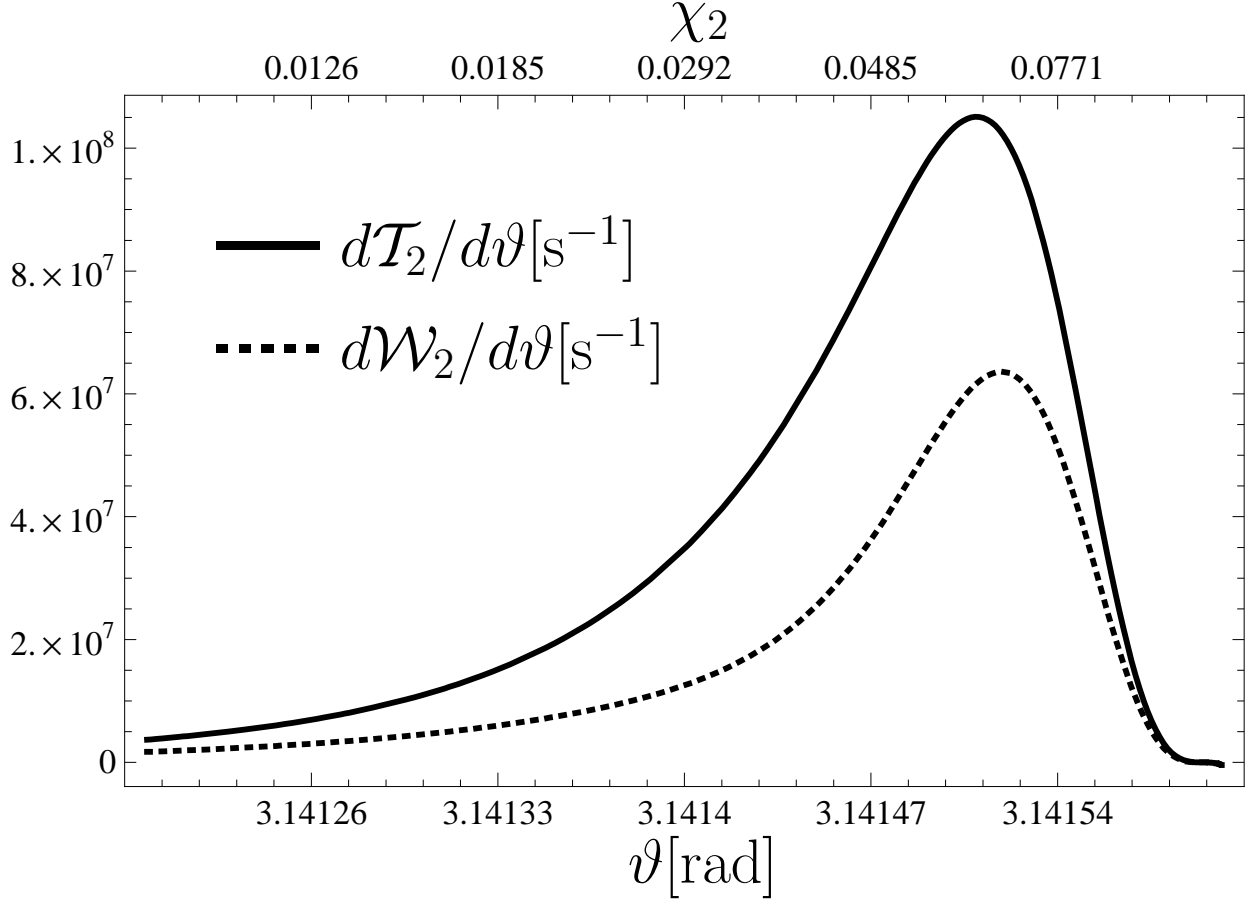}
\end{center}
\caption{The rate per unit angle $\vartheta$ of photons emitted only via 2-photon Thomson scattering (dashed line) and via both 2-photon Thomson scattering and 2-photon merging (continuous line). The upper horizontal axis shows the values of the parameter $\chi_2$ as a function of $\vartheta$. The parameters of the laser field are $\omega_0=1.55\;\text{eV}$ and $I_0=2.5\times 10^{20}\;\text{W/cm$^2$}$ and the proton's relativistic Lorentz factor is $\gamma=7\times 10^3$. }
\label{Optical_Weak}
\end{figure}

As we have mentioned in the Introduction, various proposals have already been put forward to observe perturbative VPEs in a strong laser field \cite{VPEs1,VPEs2,VPEs3,VPEs4,VPEs5,VPEs6,VPEs7,VPEs8,VPEs9,VPEs10,VPEs11,VPEs12,VPEs13,Milstein_2005}. We provide below a numerical example showing that with our scheme the requirements for the laser field to observe perturbative VPEs are easily fulfilled nowadays. As a strong laser beam we consider, in fact, a typical multiterawatt laser beam with the following parameters: a laser photon energy of $\omega_0=1.55\;\text{eV}$, a pulse energy of $5\;\text{J}$, a pulse duration of $\tau_l=25\;\text{fs}$ at $\nu_l=10\;\text{Hz}$ repetition rate and a spot radius of $\sigma_0=5\;\text{$\mu$m}$ (the resulting laser intensity is $I_0=2.5\times 10^{20}\;\text{W/cm$^2$}$) \cite{Kieffer_2006}. Due to the low photon energy we have to use a high-energy proton beam so that multiphoton Thomson scattering does not conceal the laser photon merging process. To show this, we calculate the ratio $\epsilon$ between the total rate $\mathcal{W}_2$ of 2-photon Thomson scattering \cite{Ritus_review}
\begin{equation}
\mathcal{W}_2=\frac{7\alpha}{40}(1+\beta)\omega_0\left(\frac{m^4}{\omega_0M}\right)^2\left(\frac{I_0}{I_{cr}}\right)^2,
\end{equation}
and the total rate $\mathcal{R}_2$ in Eq. (\ref{R_2}), and we obtain
\begin{equation}
\epsilon=\frac{\mathcal{W}_2}{\mathcal{R}_2}=\frac{3402\pi^2}{13\alpha^2}\left[\frac{1}{\gamma(1+\beta)}\frac{m^2}{\omega_0 M}\right]^4\approx 5\times 10^7\left[\frac{1}{\gamma(1+\beta)}\frac{m^2}{\omega_0 M}\right]^4.
\end{equation}
If we use the above laser photon energy and we assume that $\gamma\gg 1$, we find $\epsilon\approx 3.3\times 10^{15}/\gamma^4$. In this way, if we require that the background process of 2-photon Thomson scattering does not conceal our effect, i. e. that at most $\epsilon\sim 1$, then proton relativistic factors larger than $10^3$ are required that will be available at LHC. The main proton beam parameters at LHC are \cite{PDG}: a proton energy of $E_p=7\;\text{TeV}$, a number of protons per bunch of $N_p=11.5\times 10^{10}$, a bunch transversal radius of $R_p=16.6\;\text{$\mu$m}$, a bunch length of $l_p=7.55\;\text{cm}$. As we have done in \cite{Di_Piazza_2008}, in Fig. \ref{Optical_Weak} we compare the differential rate $d\mathcal{W}_2/d\vartheta$ of photons emitted only by 2-photon Thomson scattering (dashed line, result via \cite{Ritus_review}) and the differential rate $d\mathcal{T}_2/d\vartheta$ of photons emitted also by including the 2-photon merging process (continuous line). We have already observed that in general these two processes interfere because the photons emitted have the same frequency in the two cases. The contribution of the VPEs is clear from the figure and it becomes more evident if we calculate the total number of photons emitted in one hour only via 2-photon Thomson scattering and only via 2-photon merging. To do this we first integrate numerically the corresponding differential rates over the angle $\vartheta$. Then, we need the effective number of protons that pass through the laser beam in one shot, the laser pulse duration (the proton bunch is temporally much longer than the laser pulse) and then the laser repetition rate (the repetition rates of conventional proton accelerators are orders of magnitude larger than the laser repetition rate). In turn, we can estimate the effective number $N_{p,\text{eff}}$ of protons passing through the laser beam as $N_{p,\text{eff}}=N_p\times\min\{\sigma_0^2/R_p^2,1\}\times\min\{2r_l/l_p,1\}$ with $r_l=\omega_0\sigma_0^2/2$ being the laser Rayleigh length. By using these quantities we can estimate the total number of photons emitted in one hour via 2-photon merging as $3600\,N_{p,\text{eff}}\mathcal{R}_2[\text{s}^{-1}]\tau_l[\text{s}]\nu_l[\text{Hz}]$ and we obtain 180 events. Analogously, we obtain that the total number of photons emitted in one hour only via 2-photon Thomson scattering is again 180. If one includes the two processes by summing their amplitudes, one obtains about 360 events as the total number of photons produced. This shows that there is no interference between the two processes in the present regime. In fact, at small $\chi_2$ the coefficients $c_{j,2}$ are real and the amplitude of the 2-photon merging process is purely imaginary (see Eq. (\ref{M})). On the contrary, it can be shown starting from the results in \cite{Ritus_review}, that the amplitude of the 2-photon Thomson scattering is purely real. The values of the parameter $\chi_2$ shown in the upper horizontal axis in Fig. \ref{Optical_Weak} indicate that the results are perturbative in the laser field because $\chi_2\ll 1$. Also, as it is expected in the perturbative regime, the merging of 4-laser photons is completely negligible: only $5\times 10^{-4}$ events per hour. This is a clear indication of the qualitative and quantitative difference between the perturbative and the nonperturbative regime in the process of laser photon merging. It is worth observing that here the number of events per hour is not much smaller than that obtained in the numerical example shown in \cite{Di_Piazza_2008} where in the nonperturbative regime we obtained about 390 2-photon merging events per hour. This is due to the fact that here the spacetime overlapping of the laser beam and the proton beam is much larger than in \cite{Di_Piazza_2008} where a tight focused ($\sigma_0=0.8\;\text{$\mu$m}$), short ($\tau_l=5\;\text{fs}$) laser pulse was employed. In fact, the 2-photon merging rate here is about 4-5 orders of magnitude smaller than that in \cite{Di_Piazza_2008}.

In the above example a proton beam with the parameters available at the LHC is required. However, the perturbative regime of the photon merging process can be realized even in an all-optical setup using laser parameters envisaged at ELI \cite{Laser_ELI}. The proton beam could be produced in the so-called laser-piston regime by shooting an ultra-strong laser beam onto a plasma slab \cite{Esirkepov_2004}. According to the simulations of the latter, proton energies of about $E_p=50\;\text{GeV}$ can be envisaged. The other relevant proton beam parameters are: number of protons in the bunch $N_p=2\times 10^{12}$, bunch transversal radius $R_p=5\;\text{$\mu$m}$ and bunch length $l_p=6\;\text{$\mu$m}$. The relatively low proton energy can be compensated by employing a strong attosecond pulse of extreme ultraviolet (XUV) radiation (see \cite{Tsakiris_2006}): intensity of $I_0=1.4\times 10^{24}\;\text{W/cm$^2$}$, photon energy of $\omega_0=200\;\text{eV}$ and pulse duration of $\tau_l=38\;\text{as}$. This XUV pulse, according to \cite{Tsakiris_2006}, can be produced by the reflection from a plasma surface of an ultra-strong laser pulse of $5\;\text{fs}$ duration, intensity $2.5\times 10^{24}\;\text{W/cm$^2$}$, focused onto a spot-radius of $5\;\text{$\mu$m}$ which assumes conversion efficiency of $4\times 10^{-3}$ (we have extrapolated the values of the conversion efficiency that in \cite{Tsakiris_2006} are available for initial laser intensities up to about $10^{22}\;\text{W/cm$^2$}$; however, the conversion efficiency weakly depends on the laser intensity at high relativistic intensities). The dependence of the 2-photon merging rates $d\mathcal{W}_2/d\vartheta$ and $d\mathcal{T}_2/d\vartheta$ is similar to that in Fig. 5 and we don't report it. For the total rates we obtain  6.9 photons per shot only due to 2-photon Thomson scattering, 5.2 photons per shot only due to 2-laser photon merging and 12.1 photons per shot due to the two processes together.  We also stress that the simulations performed in \cite{Esirkepov_2004,Tsakiris_2006} have been carried out for laser intensities much smaller than those available at ELI. If the results of \cite{Esirkepov_2004,Tsakiris_2006} can be scaled to intensities of the order of $10^{25}\text{-}10^{26}\;\text{W/cm$^2$}$ like those available at ELI, much larger rates can be expected. Moreover, the values of the parameters $\chi_{2n}$ can reach  the order of unity 
and non-perturbative, multiphoton VPEs (merging of more than two laser photons) could become observable.
%
%
\subsubsection{Polarization and complete angular distribution of the emitted photons}

Since the present case of a strong optical laser field is the most favorable from an experimental point of view, we also investigate here for this case the polarization properties of the emitted photons and their complete angular distribution. Starting from Eq. (\ref{R}) we obtain the rate per unit of solid angle and for each polarization $a=1,2$ as
\begin{equation}
\label{R_a}
\begin{split}
\frac{d\mathcal{R}_a}{d\Omega}
&=\sum_{n=1}^{\infty}\frac{d\mathcal{R}_{2n,a}}{d\Omega}=\frac{\alpha^3}{64\pi^3}\frac{m^4}{(1+\beta)\gamma^2\omega_0^3}\sum_{n=1}^{\infty}\frac{1}{n^3}\frac{1}{(1-\cos\vartheta)^2}\left\vert\sum_{j=1}^2c_{j,2n}(u\Lambda_j)(e_a^*\Lambda_j)\right\vert^2.
\end{split}
\end{equation}
The two polarization unit vectors of the final photons have been chosen as $e_a^{\mu}=(0,\mathbf{e}_a)$ with
\begin{align}
\mathbf{e}_1&=\frac{\mathbf{x}\times\mathbf{k}}{|\mathbf{x}\times\mathbf{k}|},\\
\mathbf{e}_2&=\frac{\mathbf{k}\times\mathbf{e}_1}{|\mathbf{k}\times\mathbf{e}_1|}.
\end{align}
By employing these polarization four-vectors and the four-vectors $\Lambda^{\mu}_j$ given in Eq. (\ref{Lambda_j}), we obtain the following expressions for the four-dimensional scalar products occurring in Eq. (\ref{R_a})
\begin{align}
\label{e_1_Lambda_1}
(e_1\Lambda_1)&=\frac{1}{\sqrt{\cos^2\vartheta+\sin^2\vartheta\, \cos^2\varphi}}\left(-\cos\vartheta+\frac{\sin^2\vartheta}{1-\cos\vartheta}\cos^2\varphi\right),\\
(e_1\Lambda_2)&=(e_2\Lambda_1)=\frac{1}{\sqrt{\cos^2\vartheta+\sin^2\vartheta\, \cos^2\varphi}}\frac{\sin^2\vartheta}{1-\cos\vartheta}\sin\varphi\cos\varphi,\\
\label{e_2_Lambda_2}
(e_2\Lambda_2)&=\frac{1}{\sqrt{\cos^2\vartheta+\sin^2\vartheta\, \cos^2\varphi}}\left(-1+\frac{\sin^2\vartheta}{1-\cos\vartheta}\sin^2\varphi\right).
\end{align}
In the above paragraph we have seen that the contribution of the laser photon merging to the total rate of photons produced together with multiphoton Thomson scattering is in principle measurable. In the following we show that the contribution of the laser photon merging process is more clearly detectable by measuring the polarization and the angular distribution along the angle $\varphi$ of the final photons. To do this we reconsider the first numerical example already discussed in the above paragraph which is based on already existing laser and proton acceleration technique. In Figs. 6 and 7 we show the photon rate emitted per unit of solid angle and with polarization 1 and 2, respectively. In the part a) of the figures we have plotted only the contribution $d\mathcal{W}_{2,a}/d\Omega$ due to 2-photon Thomson scattering (see the Review \cite{Ritus_review} for a detailed discussion about multiphoton Thomson scattering of electrons) while in the part b) we have plotted the total contribution $d\mathcal{T}_{2,a}/d\Omega$ including also the 2-laser photon merging process. In the case of polarization 1 the two plots are very similar, implying that the contribution of the laser photon merging is smaller than that of 2-photon Thomson scattering. However, in the case of polarization 2 the opposite situation occurs: the contribution of the 2-laser photon merging process is dominating and the angular distribution along $\varphi$ is clearly different in the two cases. In particular, the modulation depth of the total photon rate at varying $\varphi$ and at fixed $\vartheta\approx\pi$ is at least 4 times larger than that for the Thomson scattering case only. This property will allow to distinguish experimentally the contribution of the 2-laser photon merging process without a measurement of an absolute photon number. It is also worth observing from Eqs. (\ref{e_1_Lambda_1})-(\ref{e_2_Lambda_2}) that the dependence of the two differential rates $d\mathcal{R}_{2,a}/d\Omega$ on the azimuthal angle $\varphi$ is rather complex and, in particular, it cannot be concluded that $d\mathcal{R}_{2,1}(\varphi)/d\Omega=d\mathcal{R}_{2,2}(\varphi-\varphi_0)/d\Omega$, with $\varphi_0$ being an
arbitrary fixed angle. This is a consequence of the fact that the linear polarization of the laser field along the $z$ direction breaks the cylindrical symmetry around the $y$ direction. To the sake of clarity we also present in Figs. 8 and 9 the above photon rates but integrated with respect to $\varphi$ (and multiplied times $\sin\vartheta$). The two curves in Fig. 8 corresponding to the case of polarization 1 are again very similar. However, the two curves in Fig. 9 corresponding to the case of polarization 2 are different not only because the contribution of the 2-photon Thomson scattering is very small but also because the angular distributions of the two rates show a peak at different angles $\vartheta$. In conclusion, measurements of the polarization of the final photons and of their angular distribution can be a useful experimental tool to detect the process of 2-laser photon merging in the presence of a strong optical laser field. Finally, we observe that the differential rates $d\mathcal{R}_{2,1}/d\vartheta$ and $d\mathcal{R}_{2,2}/d\vartheta$ integrated with respect to $\varphi$ can be obtained from Figs. 8 and 9 by subtracting the dotted curves from the continuous ones because the interference between 2-photon Thomson scattering and 2-photon merging is in this case negligible. As we have already mentioned, the difference between these two rates is a consequence of the symmetry breaking induced by the linear polarization of the laser field. 
%
%
\subsection{The asymptotic limit of $\eta_{2n}\gg 1$ and $\xi$ fixed}

As we have mentioned in Par. \ref{small_xi}, this asymptotic limit is of more theoretical than practical interest due to the large laser frequencies required, in general, to achieve this regime. Moreover, in the present case we assume that the parameter $\xi$ is not much smaller than unity and this circumstance is verified only by employing the rather speculative ``Goal'' parameters for the future X-FEL at DESY as given in \cite{Ringwald_2001}. However, this regime is interesting because it can indicate modifications of the behavior of QED at high energies due to the presence of a strong laser field. This circumstance has already been noted, for example, in \cite{Milstein_2006}, in investigating the pair-creation process in a strong laser field and a nuclear field. Whereas, the opposite situation was found in \cite{Di_Piazza_2007_a} where at high energies of the incoming photon the photon splitting amplitude resulted the same as that of photon-photon scattering in vacuum.

In order to find the leading term of the coefficients $c_{j,2n}$ it is more convenient to take first the integral on $v$ in Eq. (\ref{c_j_n_ex}). This integral can be performed from $0$ to $1$ due to the symmetry of the integrand and then it is convenient to change the variable according to: $s=1-v$. Now, in the asymptotic limit $\eta_{2n}\gg 1$ it is possible for each $n$ to find a constant $\varepsilon_{2n}$ such that $\eta_{2n}^{-1}\ll \varepsilon_{2n}\ll 1$. After that, one divides the integral over $s$ into two integrals, one from $0$ to $\varepsilon_{2n}$ and one from $\varepsilon_{2n}$ to $1$. In this way the coefficients $c_{j,2n}$ receive two contributions and they can be written as $c_{j,2n}=c^{<}_{j,2n}(\varepsilon_{2n})+c^{>}_{j,2n}(\varepsilon_{2n})$ for which we have pointed out that the two single contributions $c^{<}_{j,2n}(\varepsilon_{2n})$ and $c^{>}_{j,2n}(\varepsilon_{2n})$ (but not their sum) depend, in general, on the arbitrary quantity $\varepsilon_{2n}$. The calculation of the leading terms in $\eta_{2n}^{-1}$ to $c^{<}_{j,2n}(\varepsilon_{2n})$ and $c^{>}_{j,2n}(\varepsilon_{2n})$ is lengthy but straightforward. Below, we give the final results:
\begin{equation}
\label{c_j_2_large_eta}
\begin{split}
c_{j,2}&\sim \int_0^{\infty}\frac{d\rho}{\rho} e^{-2i\rho}\left\{\xi^2\sin^{2(j-1)}(\rho)\left(\frac{b}{a}\right)\left(1+\sqrt{1-\frac{b^2}{a^2}}\right)^{-1}\right.\\
&\left.+\frac{\xi^2}{2}\sin^2\rho\left[\log\left(1+\sqrt{1-\frac{b^2}{a^2}}\right)+C+i\frac{\pi}{2}-\log\left(\frac{4\eta_2}{a}\right)-\frac{b^2}{2a^2}\left(1+\sqrt{1-\frac{b^2}{a^2}}\right)^{-2}\right]\right.\\
&+i\frac{b^3}{16a^2}{_3F_2}\left(1,1,\frac{3}{2};2,3;\frac{b^2}{a^2}\right)+\frac{\xi^2}{2}\left(1+\frac{\sin^2\rho}{\rho^2}-\frac{\sin 2\rho}{\rho}\right)(2-j)+\\
&-i\frac{b}{2}\left(C+i\frac{\pi}{2}-\log\left(\frac{2\eta_2}{a}\right)+\frac{i}{2\rho}\right)\Bigg\}
\end{split}
\end{equation}
if $n=1$, and
\begin{equation}
\label{c_j_2n_large_eta}
\begin{split}
c_{j,2n}&\sim 2i^n\int_0^{\infty}\frac{d\rho}{\rho} e^{-2in\rho}\left\{\frac{\xi^2}{2n}\sin^{2(j-1)}(\rho)\left(\frac{b}{ia}\right)^n\left(1+\sqrt{1-\frac{b^2}{a^2}}\right)^{-n}\right.\\
&+i\frac{\xi^2}{4}\sin^2\rho\left[\frac{1}{n-1}\left(\frac{b}{ia}\right)^{n-1}\left(1+\sqrt{1-\frac{b^2}{a^2}}\right)^{-(n-1)}\right.\\
&\left.-\frac{1}{n+1}\left(\frac{b}{ia}\right)^{n+1}\left(1+\sqrt{1-\frac{b^2}{a^2}}\right)^{-(n+1)}\right]\\
&\left.+\frac{b}{2(n^2-1)}\left(\frac{b}{ia}\right)^{n-1}\left(1+n\sqrt{1-\frac{b^2}{a^2}}\right)\left(1+\sqrt{1-\frac{b^2}{a^2}}\right)^{-n}\right\}
\end{split}
\end{equation}
if $n>1$. In Eq. (\ref{c_j_2_large_eta}) we have introduced the Euler constant $C=0.577216\ldots$ and the hypergeometric function $_3F_2(\alpha_1,\alpha_2,\alpha_3;\beta_1,\beta_2;z)$, with $\alpha_1,\alpha_2,\alpha_3$ and $\beta_1,\beta_2$ being arbitrary real parameters and with $z$ being in general a complex variable \cite{Gradshteyn_b_2000}. We have also used the following notation
\begin{align}
a&=2\rho\left[1+\frac{\xi^2}{2}\left(1-\frac{\sin^2\rho}{\rho^2}\right)\right],\\
b&=\rho\xi^2\left(\frac{\sin^2\rho}{\rho^2}-\frac{\sin 2\rho}{2\rho}\right).
\end{align}
It is worth noting that in the present case the amplitudes have a complex dependence on the two parameters $\xi$ and $\eta_2$ \emph{independently}, while in the case of an optical laser field they depend only on the combination $\chi_{2n}=\xi\eta_{2n}$. At large $\eta_2$ the dominant contribution with logarithmic accuracy comes from Eq. (\ref{c_j_2_large_eta}) and it is given by
\begin{equation}
c_{j,2}\sim \frac{\xi^2}{4}\log 2\;\log\eta_2.
\end{equation}
By employing the numerical parameters indicated as ``Goal'' in [28], i. e. $\omega_0=8.3\;\text{keV}$ and $I_0=2.3\times 10^{27}\;\text{W/cm$^2$}$, we obtain $\xi\approx 10$ and $\eta_2\approx 4220$ at $\vartheta\approx \pi$ and $\gamma=1000$ and then $c_{j,2}\approx 145$. However, the final number of photon merging events per hour turns out to be negligibly small due to the fact that the laser field is supposed to be focused to one wavelength of $0.15\;\text{nm}$ to obtain the quoted intensity. Then the overlapping volume between laser and proton beams is very small.

We have checked that if $\xi\ll 1$ then the largest coefficients are those with $n=1$ and their expressions coincide with those already found in Par. III A (see Eq. (\ref{c_j_n_xi_small_large_eta})). In the strong-field limit $\xi\gg 1$ we observe that both $a$ and $b$ become proportional to $\xi^2$. Then, independently of the number of laser photons that merge the amplitudes result proportional to $\xi^2$. Moreover, the coefficients $c_{j,2}$ also maintain a nontrivial logarithmic dependence on $\xi$ through the function $\log(\eta_2/a)$. It is interesting to note that in this limit the amplitudes $c_{j,2}$ depend effectively on the parameter $\eta^*_2=(k_0k_2)/m^{*2}$, with $m^*=m\sqrt{1+\xi^2}\approx m\xi$ being the electron mass dressed in the laser field.
%
%
\section{Summary and conclusions}
In conclusion, in this paper we have continued the study of the process of laser photon merging in laser-proton collisions that we started in \cite{Di_Piazza_2008}. We confirm that the most interesting regime from an experimental point of view is that investigated in \cite{Di_Piazza_2008}, which offers the possibility of observing at least in principle nonperturbative refractive vacuum polarization effects (VPEs) in a laser field. In fact, we have seen that  the  higher intensity of strong available optical laser fields compensates for the high photon energy, the high repetition rate of an X-FEL and for the large overlapping region between the X-FEL and the proton beams. We have shown here with a numerical example that, in order to observe perturbative VPEs with our setup, optical laser intensities of the order of $10^{20}\;\text{W/cm$^2$}$ are already sufficient. Due to the experimental relevance of the case of a strong optical laser field, we have also investigated for this case the polarization and the complete angular distribution of the emitted photons. Interestingly, we have seen that both polarization measurements and angular distribution measurement can be performed to clearly discriminate between the process of laser photon merging and the background process of multiphoton Thomson scattering.

The other regimes of parameters studied in the present paper have shown interesting features as well. First, we have found the leading-order contribution to the coefficients $c_{j,2n}$ for small $\xi$ and arbitrary $\eta_{2n}$. We have seen that this regime is suitable when the laser beam used to polarize the vacuum is an X-FEL. The leading-order contributions in this regime are those corresponding to 2-laser photon merging and they are expressed only through elementary functions of $\eta_2$ without any further integration. Moreover, we have studied the limit of large $\eta_{2n}$ and fixed $\xi$ and we have seen that the coefficients $c_{j,2n}$ show a complex dependence on the relativistic parameter $\xi$. In the strong field limit $\xi\gg 1$ the amplitudes show the same proportionality to $\xi^2$, independently of the number of laser photons that merge. Finally, in the same limit the coefficients $c_{j,2}$ (corresponding to the 2-photon merging process) also show a nontrivial logarithmic dependence on $\xi$.
\clearpage
%
%
\appendix

\section{}

We report here the final expression found in \cite{Baier_1976} for the polarization operator in the plane wave with four-vector potential
\begin{equation}
\label{A_mu}
A^{\mu}(x)=A_{10}a_1^{\mu}\cos(k_0x)+A_{20}a_2^{\mu}\sin(k_0x),
\end{equation}
where $a_1^{\mu}=(0,0,0,1)$, $a_2^{\mu}=(0,1,0,0)$ and where $k_0^{\mu}=(\omega_0,0,\omega_0,0)$ is the four-momentum of the laser photons. We consider only the case where the initial photon with four-momentum $q^{\mu}$ is virtual \textit{and $q^2<0$}, while the final one with four-momentum $k^{\mu}$ is real and $k^2=0$. The polarization operator $\Pi^{\mu\nu}(q,k)$ can be expressed as (we note that in the more general case with $k^2\neq 0$ a fifth term would be present in the following sum)
\begin{equation}
\Pi^{\mu\nu}(q,k)=C_1(q,k)\Lambda_1^{\mu}\Lambda_2^{\nu}+C_2(q,k)\Lambda_2^{\mu}\Lambda_1^{\nu}+C_3(q,k)\Lambda_1^{\mu}\Lambda_1^{\nu}+C_4(q,k)\Lambda_2^{\mu}\Lambda_2^{\nu}.
\end{equation}
In this expression the four-vectors $\Lambda_j^{\mu}$ with $j\in \{1,2\}$ are given by
\begin{equation}
\label{Lambda_j_app}
\Lambda_j^{\mu}=a_j^{\mu}-\frac{(ka_j)}{(k_0k)}k_0^{\mu},
\end{equation}
and the coefficients $C_r(q,k)$ with $r\in\{1,\ldots, 4\}$ can be written as
\begin{equation}
\label{C_r}
\begin{split}
C_r(q,k)&=-i(2\pi)^4m^2\frac{\alpha}{\pi}\int_{-1}^1 dv\int_0^{\infty}\frac{d\rho}{\rho}\exp\left\{-i\frac{2\rho}{\lambda(1-v^2)}\left[1-\frac{(kq)}{4m^2}(1-v^2)+\mathcal{A}(\xi_1^2+\xi_2^2)\right]\right\}\\
&\times\sum_{l=1}^{\infty}\delta(q+2lk_0-k)g_{r,l},
\end{split}
\end{equation}
where $\delta(p)$ is the four-dimensional Dirac $\delta$ function and where
\begin{align}
g_{1,l}&=\xi_1\xi_2\left(2\mathcal{A}_0\rho\frac{1+v^2}{1-v^2}+\mathcal{A}_1\frac{l}{z}\right)i^{-l}J_{-l}(z),\\
g_{2,l}&=\xi_1\xi_2\left(-2\mathcal{A}_0\rho\frac{1+v^2}{1-v^2}+\mathcal{A}_1\frac{l}{z}\right)i^{-l}J_{-l}(z),\\
\begin{split}
g_{3,l}&=\left(\xi_1^2\mathcal{A}_1+\frac{\xi_1^2v^2+\xi_2^2}{1-v^2}\sin^2\rho\right)i^{-l}J_{-l}(z)+\left(\xi_1^2\mathcal{A}_1-\frac{\xi_1^2-\xi_2^2}{1-v^2}\sin^2\rho\right)i^{-l-1}J'_{-l}(z)\\
&-\frac{1}{4}\left[\frac{(kq)}{m^2}+i\frac{\lambda(1-v^2)}{\rho}\right]i^{-l}J_{-l}(z),
\end{split}\\
g_{4,l}&=(-1)^lg_{3,l}(\xi_1^2\longleftrightarrow\xi_2^2),
\end{align}
with $\xi_j^2=-e^2A_{j0}^2/m^2$, $J_n(z)$ being the ordinary Bessel function of order $n$ and $J'_n(z)$ its derivative and with
\begin{align}
\mathcal{A}&=\frac{1}{2}\left(1-\frac{\sin^2\rho}{\rho^2}\right),\\
\mathcal{A}_0&=\frac{1}{2}\left(\frac{\sin^2\rho}{\rho^2}-\frac{\sin 2\rho}{2\rho}\right),\\
\mathcal{A}_1&=\mathcal{A}+2\mathcal{A}_0,\\
z&=\frac{2\rho(\xi_1^2-\xi_2^2)}{\lambda(1-v^2)}\mathcal{A}_0,\\
\lambda&=\frac{(k_0k)}{2m^2}.
\end{align}
\textit{We mention that, in general, the series in Eq. (\ref{C_r}) also contains terms with $l\le 0$. However, their contributions vanish in our case because the corresponding $\delta$ functions cannot be fulfilled. In fact, from the relation $q^\mu+2lk^\mu_0-k^\mu=0$ it follows that $q^2=-4l(k_0k)<0$. Then, since $(k_0k)>0$, it must be $l>0$.}
%
%
\begin{figure}
\begin{center}
\includegraphics[width=\textwidth]{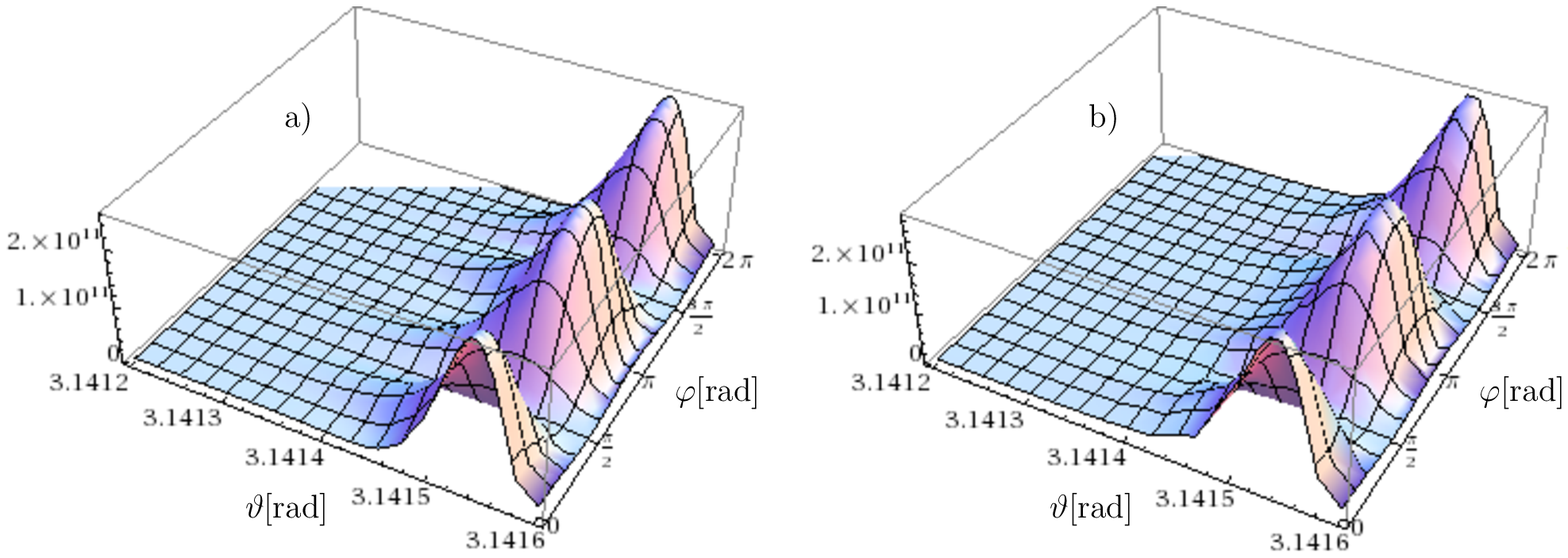}
\end{center}
\caption{(Color online) The rate in sec$^{-1}$ per unit of solid angle $d\Omega=\sin\vartheta d\vartheta d\varphi$ of photons emitted per second with polarization 1 only via 2-photon Thomson scattering (part a)) and via both 2-photon Thomson scattering and 2-laser photon merging (part b)). The parameters of the laser field and of the proton are the same as in Fig. 5.}
\label{Optical_Weak_dOmega_Pol_1}
\end{figure}
\begin{figure}
\begin{center}
\includegraphics[width=\textwidth]{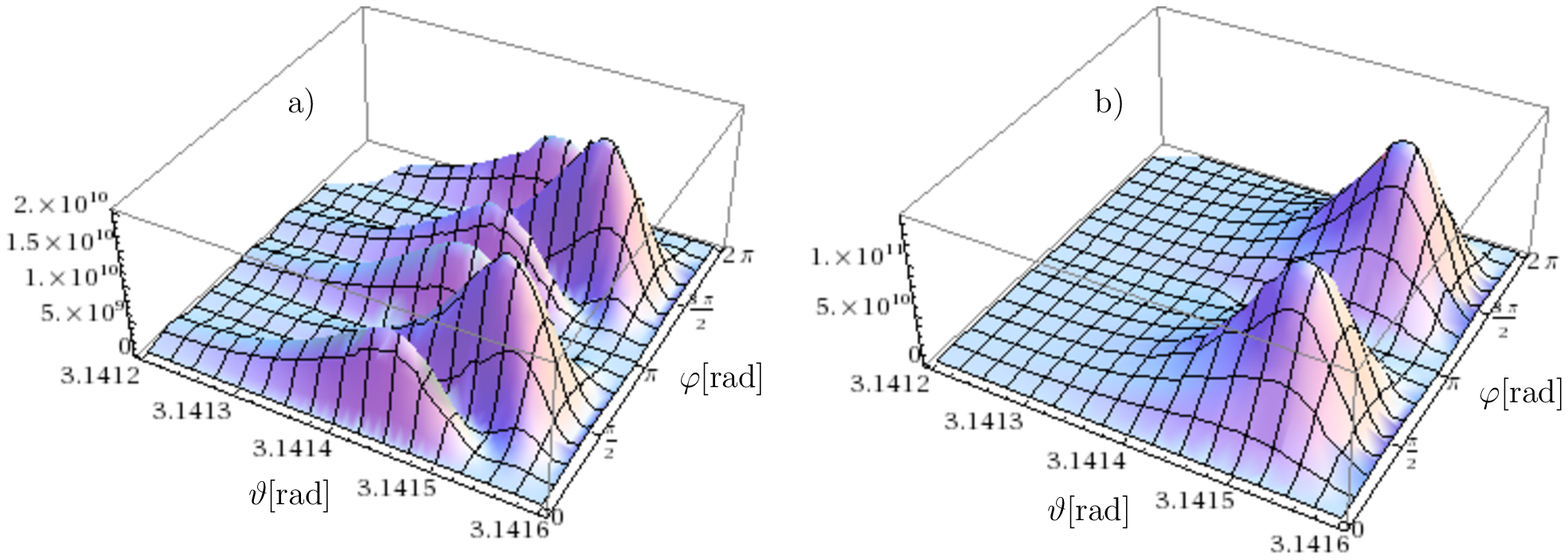}
\end{center}
\caption{(Color online) The rate in sec$^{-1}$ per unit of solid angle $d\Omega=\sin\vartheta d\vartheta d\varphi$ of photons emitted with polarization 2 only via 2-photon Thomson scattering (part a)) and via both 2-photon Thomson scattering and 2-laser photon merging (part b)). The parameters of the laser field and of the proton are the same as in Fig. 5.}
\label{Optical_Weak_dOmega_Pol_2}
\end{figure}
\begin{figure}
\begin{center}
\includegraphics[width=\textwidth]{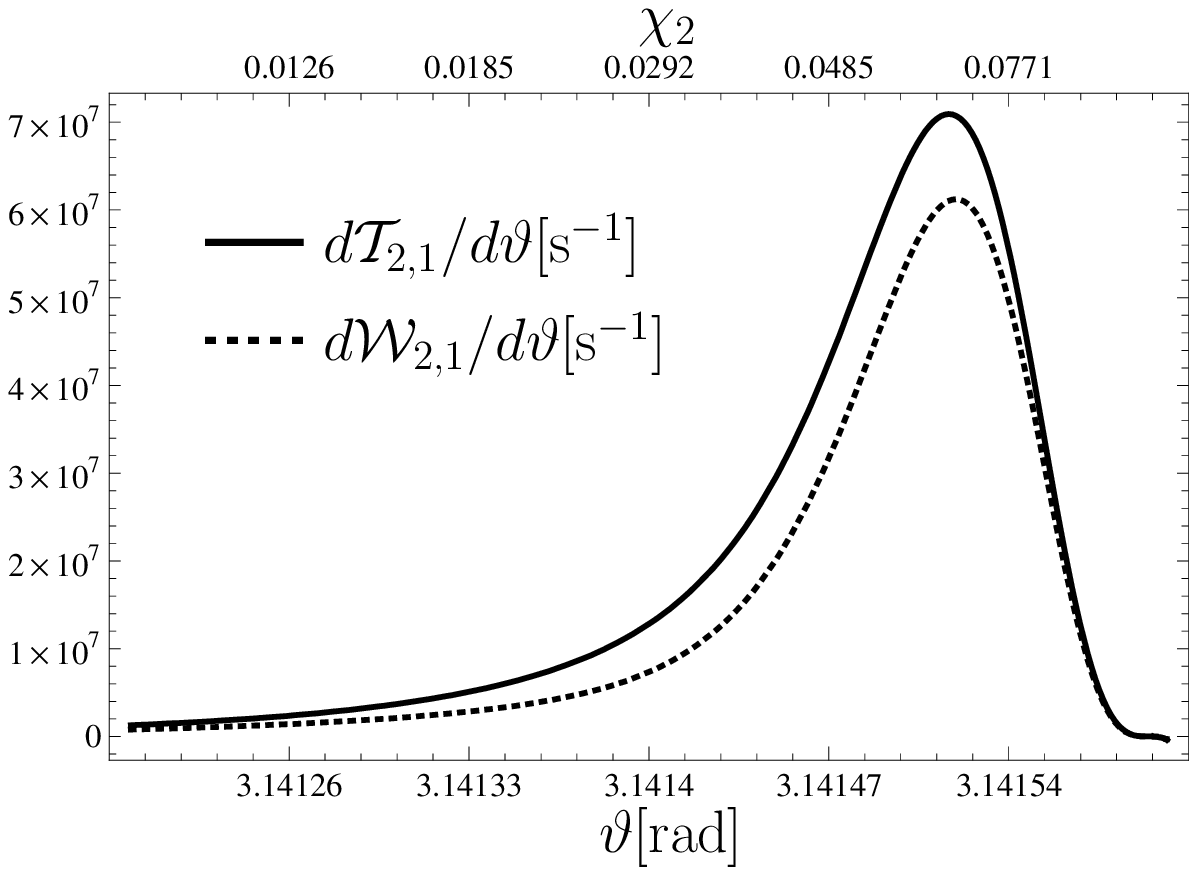}
\end{center}
\caption{The rate per unit angle $\vartheta$ of photons emitted with polarization 1 only via 2-photon Thomson scattering (dashed line) and via both 2-photon Thomson scattering and 2-photon merging (continuous line). The upper horizontal axis shows the values of the parameter $\chi_2$ as a function of $\vartheta$. The parameters of the laser field and of the proton are the same as in Fig. 5.}
\end{figure}
\begin{figure}
\begin{center}
\includegraphics[width=\textwidth]{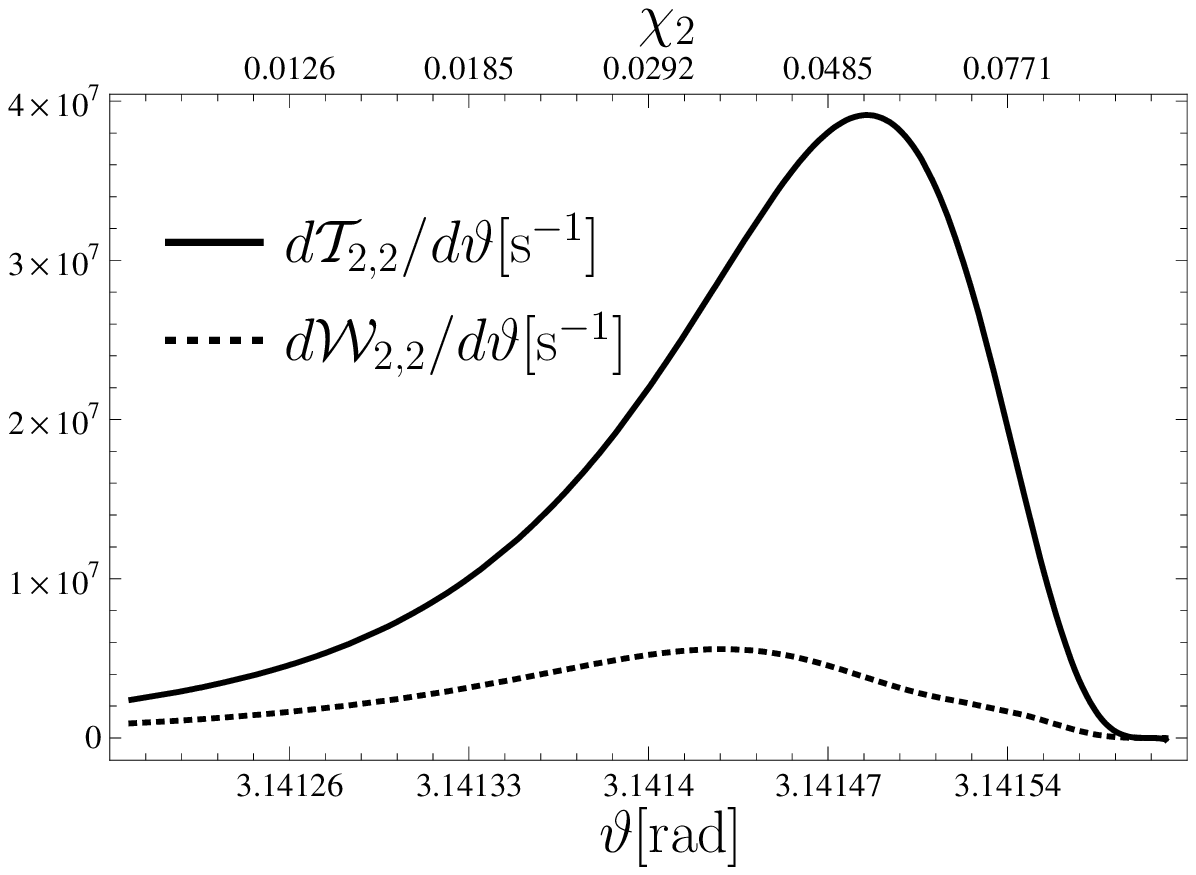}
\end{center}
\caption{The rate per unit angle $\vartheta$ of photons emitted with polarization 2 only via 2-photon Thomson scattering (dashed line) and via both 2-photon Thomson scattering and 2-photon merging (continuous line). The upper horizontal axis shows the values of the parameter $\chi_2$ as a function of $\vartheta$. The parameters of the laser field and of the proton are the same as in Fig. 5.}
\end{figure}
\clearpage
%
%
%

\end{document}